\documentclass[letter]{aa}
\usepackage{graphicx}
\usepackage{url}

\usepackage{txfonts}
\usepackage{natbib}
\newcommand{\shen}{Soph to be submitted}

\newcommand{\speed}[1]{#1 km~s${}^{-1}$}

\begin{document}

\title{Total reflection of a flare-driven quasi-periodic EUV wave train at a coronal hole boundary}
\author{Xinping Zhou\inst{1,2}, Yuandeng Shen\inst{1,2}, Zehao Tang\inst{1,2}, Chengrui Zhou\inst{1,2}, Yadan Duan\inst{1,2}, Song Tang\inst{1,2}}

\institute{
${^1}$ Yunnan Observatories, Chinese Academy of Sciences,  Kunming, 650216, China \\ 
${^2}$  University of Chinese Academy of Sciences, Beijing, China \\ \email{ydshen@ynao.ac.cn}}
\date{Received / Accepted}

\abstract{
The reflection, refraction, and transmission of large-scale extreme ultraviolet (EUV) waves (collectively, secondary waves) have been observed during their interactions with coronal structures such as active regions (ARs) and coronal holes (CHs). However, the effect of the total reflection of EUV waves has not been reported in the literature. Here, we present the first unambiguous observational evidence of the total reflection of a quasi-periodic EUV wave train during its interaction with a polar CH. The event occurred in NOAA AR 12473, located close to the southeast limb of the solar disk, and was characterized by a jet-like CME. In this study, we focus in particular on the driving mechanism s of the quasi-periodic wave train and the total reflection effect at the CH boundary. We find that the periods of the incident and the reflected wave trains are both about 100 seconds. The excitation of the quasi-periodic wave train was possibly due to the intermittent energy release in the associated flare since its period is similar to that of the quasi-periodic pulsations in the associated flare. Our observational results showed that the reflection of the wave train at the boundary of the CH was a total reflection because the measured incidence and critical angles satisfy the theory of total reflection, i.e., the incidence angle is less than the critical angle.

}
\titlerunning{Total reflection of a flare-driven quasi-periodic EUV wave train at a coronal hole boundary}
\authorrunning{Zhou et al.}
\maketitle

\section{Introduction}
As a fundamental physical phenomenon, waves of any kind can infer the physical parameters of the medium supporting them and hence provide a condition for diagnosing coronal physical parameters such as the magnetic field strength through seismology diagnostics techniques. The coronal seismology proposed by \cite{1970PASJ...22..341U,1984ApJ...279..857R} is one of the most critical techniques to estimate the magnetic--field strength of the solar corona, especially in the absence of direct approaches. This technique, based on the MHD waves and the oscillations in the corona, has been widely used in many articles \citep[e.g.,][]{2011ApJ...736L..13L,2012ApJ...753...52L,2019ApJ...873...22S,2021SoPh..296..169Z}.

Coronal waves, a ubiquitous phenomenon in the solar atmosphere, was firstly discovered by the Extreme-Ultraviolet (EUV) Imaging Telescope \citep[EIT:][]{1995SoPh..162..291D} onboard the {\em Solar and Heliospheric Observatory} \citep[SOHO:][]{1995SoPh..162....1D} and always accompanied with energetic eruptions such as flares, coronal mass ejections (CMEs) \citep[e.g.,][]{2000SoPh..196..157V,2000SoPh..196..181V}. In on-disk observations, these disturbances always appear as a circular or semicircular wavefront globally propagating through the solar corona \citep{1997SoPh..175..571M,1998GeoRL..25.2465T}. They often have a lifetime ranging from 2 to 70 minutes and are characterized by radial propagating motion with speeds of \speed{200-1500} \citep{2013ApJ...776...58N}. These EUV waves are strongly considered to be driven by the CME. Nonetheless, without a related CME, some observations indicated that there are some EUV waves closely associated with the flares \citep{2016ApJ...828...28K,2015ApJ...803L..23K}. Early observations indicated that most of the EUV waves are single-pulse, diffuse wavefronts extending across a significant fraction of the solar corona. However, more detailed information on these waves have been detected by the high spatio-temporal resolution Atmospheric Imaging Assembly \citep[AIA:][]{2012SoPh..275...17L} telescope onboard Solar Dynamics Observatory \citep[SDO:][]{2012SoPh..275....3P}. These new features, such as the bimodal composition with both wave and non-wave components \citep{2004A&A...427..705Z,2012SoPh..281..187P,2017ApJ...834L..15Z}, let their physical interpretations to a converged view, i.e., a fast-mode magnetosonic wave component travels ahead of a non-wave component. These bimodal compositions have been predicted in the 2D MHD model \citep[e.g.,][]{2002ApJ...572L..99C}. The 3D global MHD simulations further confirm this bimodal composition of an outer, faster-mode magnetosonic wave component and an inner, CME compression front \citep[e.g.,][]{2009ApJ...705..587C,2012ApJ...750..134D}. More convincing evidence comes from the off-limb events that a faster EUV wavefront separates from a CME flank and propagate freely on the solar surface once the CME has propagated sufficiently away from the Sun \citep{2011ApJ...738..160M,2011ApJ...732L..20C,2012ApJ...752L..23S,2013ApJ...773L..33S,2014ApJ...786..151S,2021SoPh..296..169Z}, resulting in a bifurcation structure in time-distance stack plots.

Recently, a new type of EUV wave train with multiple wavefronts propagating across the solar surface has been reported only in sporadic cases \citep{2012ApJ...745L..21L,2017ApJ...844..149K,2019ApJ...873...22S,2021SoPh..296..169Z}. According to the statistic by \cite{shen2021sorev}, these EUV wave trains travel away from the flare kernel with speeds ranging from \speed{370--1100} and with a period in the range of 36--240 seconds. Considering that the amplitude and velocity of these waves are not within the range of  quasi-periodic fast propagating (QFP) wave trains channeled in the open- or closed-loops but are similar to the classical EUV wave, thus \cite{shen2021sorev} classify these wave trains as broad wave trains and QFP wave trains as a narrow wave train. Otherwise, they can excite the oscillation of the cavity and filament, which are also widely observed in the classical EUV waves.  About the driving mechanism of the quasi-periodic EUV wave train, there are several possible interpretations, such as driven by CME downward and lateral compression \citep{2012ApJ...745L..21L} or unwinding motion of the filament helical structures \citep{2019ApJ...873...22S}. This divergence may be caused by the limitation of the observational case resulting in that it is hard to study the driving mechanism s detailedly and comprehensively. For detailed discussions about these two types of wave trains, we refer to the recent review by \cite{shen2021sorev}

The true wave characteristics, such as refractions, reflections, and transmissions, have manifested in the observations and simulations when a wave interact with the region exhibiting a sudden density drop, such as CHs \citep{2006ApJ...647.1466V,2010ApJ...713.1008S,2012ApJ...754....7S,2012ApJ...756..143O,2016ApJ...828...28K,2017ApJ...844..149K,2018ApJ...864L..24L} and ARs \citep{ 2002ApJ...574..440O,2013ApJ...773L..33S,2019ApJ...871L...2M}. The reflection evidence of an EUV wave at the boundary even have been observed in the radio dynamic spectra \citep{2021A&A...651L..14M}. However, the total reflection, a basic phenomenon in wave theory, in an EUV wave has not yet been identified or detected. Recently, \cite{2021A&A...651A..67P} used an extended theoretical approach, based on a linear theory treating EUV waves as fast-mode MHD waves, to investigate geometrical properties of secondary waves caused by the interaction between the oblique incoming waves and CHs. Their results show that the second waves' geometrical properties depend on the incidence angle of the incoming waves to the CH boundaries and the plasma density contrasts $\rho_c$ between the inside and outside CHs. It allows us to determine whether the secondary waves are partial or total reflected waves at the CH boundaries. In this letter, for the first time, we report the direct observational evidence of the total reflection that occurred on 2015 December 22, in which an EUV wave train emanated from the flare kernel and was totally reflected by a remote southern polar CH. The present study focus on the driving mechanism  of the EUV wave train, its propagation, and its interaction with CH. The used observational data are described in Section 2; results are presented in Section 3; discussions and conclusions are given in Section 4.

\section{Observations and data Reduction} 
The present event was recorded by the Atmospheric Imaging Assembly \cite[AIA;][]{2012SoPh..275...17L} onboard the {\em Solar Dynamic Observatory} (SDO). During our observing time interval, AIA/SDO provided continuous full-disk observations of the solar chromosphere and corona in seven extreme ultraviolet (EUV) channels, spanning a temperature range from approximately $2 \times 10^{4}$ Kelvin to over $2 \times 10^{7}$ Kelvin. Here, we mainly use the 171 \AA\ (Fe {\sc ix}; characteristic temperature: $0.6 \times 10^{6}$ K), 193 \AA\ (Fe {\sc xii}, {\sc xxiv}; characteristic temperature: $1.6\times10^{6}, 2 \times10^{7} $K) and 211 \AA\ (Fe {\sc xiv}; characteristic temperature: $2 \times 10^{6}$ K) images, since the evolutional processes are observed in these three cooler passbands, but is completely absent in the other AIA channels. The time cadence and pixel size of AIA images are 12 second and 0.6\arcsec, respectively. All of the AIA images used here were calibrated with the standard procedure available in the SolarSoftWare package provided by the instrumental team. The Geostationary Operational Environmental Satellite (GOES) soft X-ray 0.5--4.0 \AA\ and 1.0--8.0 \AA\ flux are used to analyze the periodicity pulsation of the flare. In addition, Large Angle and Spectrometric COronagraph \citep[LASCO;][]{1995SoPh..162..357B} images are used to portray the associated CME. To enhance the moving feature of the wave front, we utilized the running-difference images (i.e., each image is subtracted by the previous one) to study the wave trains' evolution. To visualize the obtained signatures of the wave train, we performed the wavelet analysis \citep{1998BAMS...79...61T} method to investigate the periodicity of the wave train and flare. We perform the Differential Emission Measure (DEM) analysis using the inversion code developed by \cite{2012A&A...539A.146H} to estimated the plasma density inside and outside CH. The DEM inversion was done at 03:22 UT before the incident wave trains arrived the CH boundary ten minutes. Additionally, we used the Collection of Analysis Tools for Coronal Holes (CATCH) algorithm developed by \cite{2019SoPh..294..144H} to extract the CH boundary, which applies the intensity gradient along the CH boundary to modulated the extraction threshold.

\section{Results} 
The event occurred close to the eastern solar limb on 2015 December 22 from the NOAA AR 12473, which associated with an M 1.6 solar flare, whose start, peak, and end times were at about 03:15 UT, 03:34 UT, and 03:48 UT, respectively (see Fig.~\ref{overview}(e)). In addition, a jet-like CME with an average speed of \speed{260} is observed by the LASCO/C2 during the eruption (see Fig.~\ref{overview}(d)).\footnote[1]{\url{http://cdaw.gsfc.nasa.gov/CME_list} }An overview of the eruption source region is presented in Fig.~\ref{overview}(a) and (b), in which there was a large southern polar CH, whose boundary determined by using the CATCH soft are also indicated in Fig.~\ref{evolution} and  Fig.~\ref{dem} (a) and (b1). The CHs regions always manifest as dark structures in the EUV and X-ray emission in comparison to the peripheral solar corona because of the reduction of density and temperature caused by plasma depletion in these regions \citep[e.g.,][]{2019SoPh..294..144H}. During  the rising phase of the flare, a EUV wave train, significantly different from the classic EIT wave that exhibits as a single large-scale quasi-circular propagating front, emanated from the flare kernel and  interacted with the distant coronal hole. More distinction between the EUV and EIT waves see a discussion in the reviews by \cite{2011JASTP..73.1096Z} and \cite{2014SoPh..289.3233L}. Recent investigation indicated that the temperature distribution of the CH has a dominant component centered around 0.9 MK and a secondary smaller component at 1.5--2.0 MK \citep{2020SoPh..295....6S}. In the AR, there exited some closed and open loops as shown in Fig.~\ref{overview}(c).

\subsection{Kinematics of the Large-Scale quasi-periodic EUV Wave train}
\begin{figure*}
\centering
\includegraphics[width=15cm]{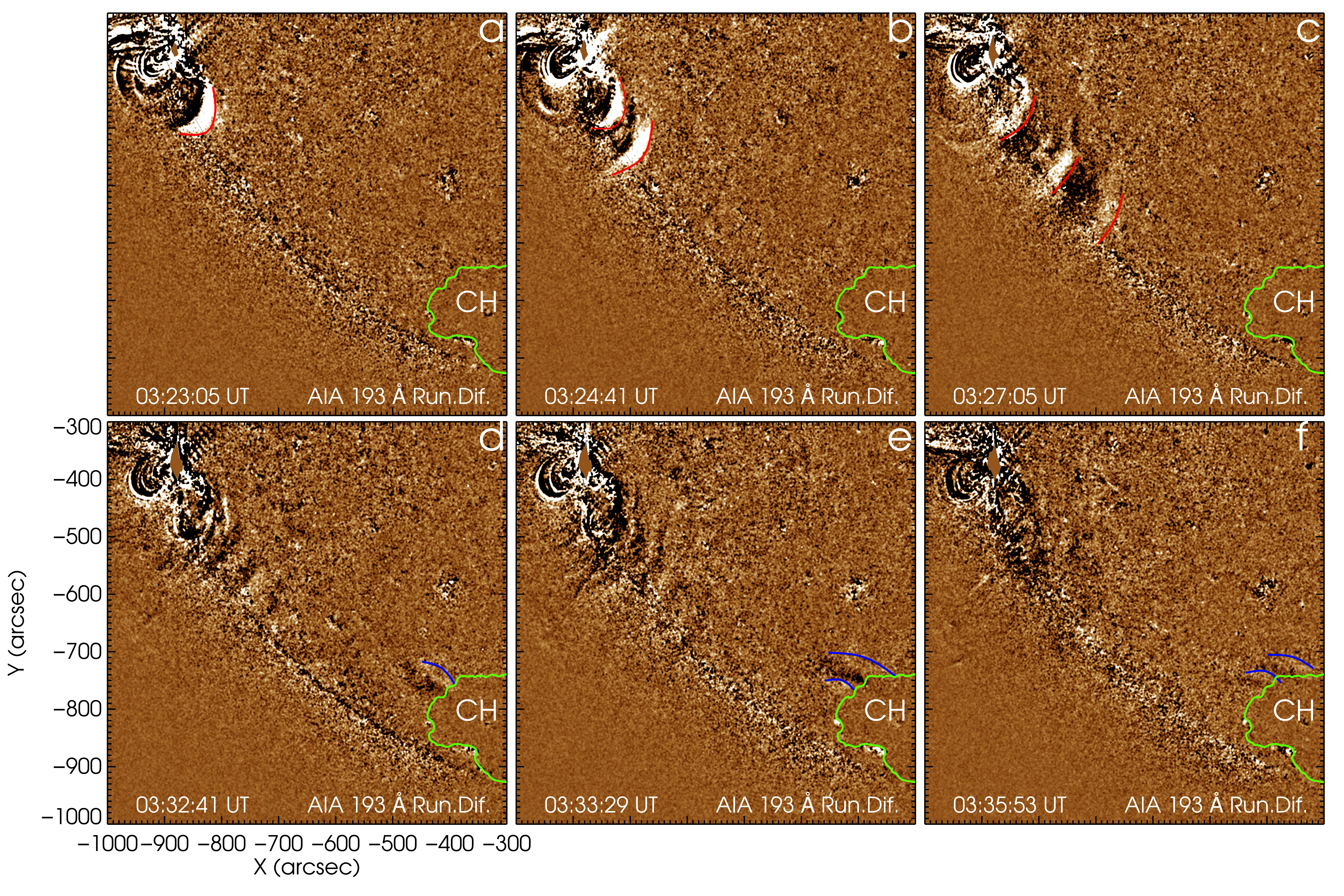}
\caption{The top and bottom rows show the evolutions of the incident and reflected wave trains in 193 \AA\ running-difference images, respectively. In each panel the red and blue curve lines depict the wavefronts of the incident and reflected EUV wave trains, respectively, while the green lines mark the location of the CH boundary.}
\label{evolution}
\end{figure*}

We mainly analyzed the time sequence of 193 \AA\ running-difference images taken during the flare to show the overall evolutionary and interactional process, since the evolutionary process are similar in 171 \AA\ and 211 \AA\ images (see the animation available in the Electronic Supplementary Material). The violent eruption launched multiple striking arc-shaped wavefronts running along the solar surface. Fig.~\ref{evolution} displays the selected running-difference images in those 193 \AA\ images showing the multiple successive wavefronts during the impulsive phase. The first wavefront appeared at about 03:22 UT, which was about 7 minutes delay the onset of the flare. Following the first wavefront, at least three successive wavefronts have been detected and these wavefronts are also clearly seen in 171 \AA\ and 211 \AA\ running-difference images, suggesting that the responsible plasma was in the range of this three channels' peak response temperature of 0.6--2.0 MK. If we carefully observe the animation and Fig.~\ref{evolution}, we can find that the intensity of the wavefronts progressively decreases as the propagation distance increases. After arriving at the CH boundary, they showed a significant reflection feature. The deflection is with respect to the initial propagation direction (and it would be around $90^\circ$ indeed, see the animation available in the Electronic Supplementary Material and Fig.~\ref{evolution} (d)--(f)). 

\begin{figure}
    \centering
    \includegraphics[width=7.5cm]{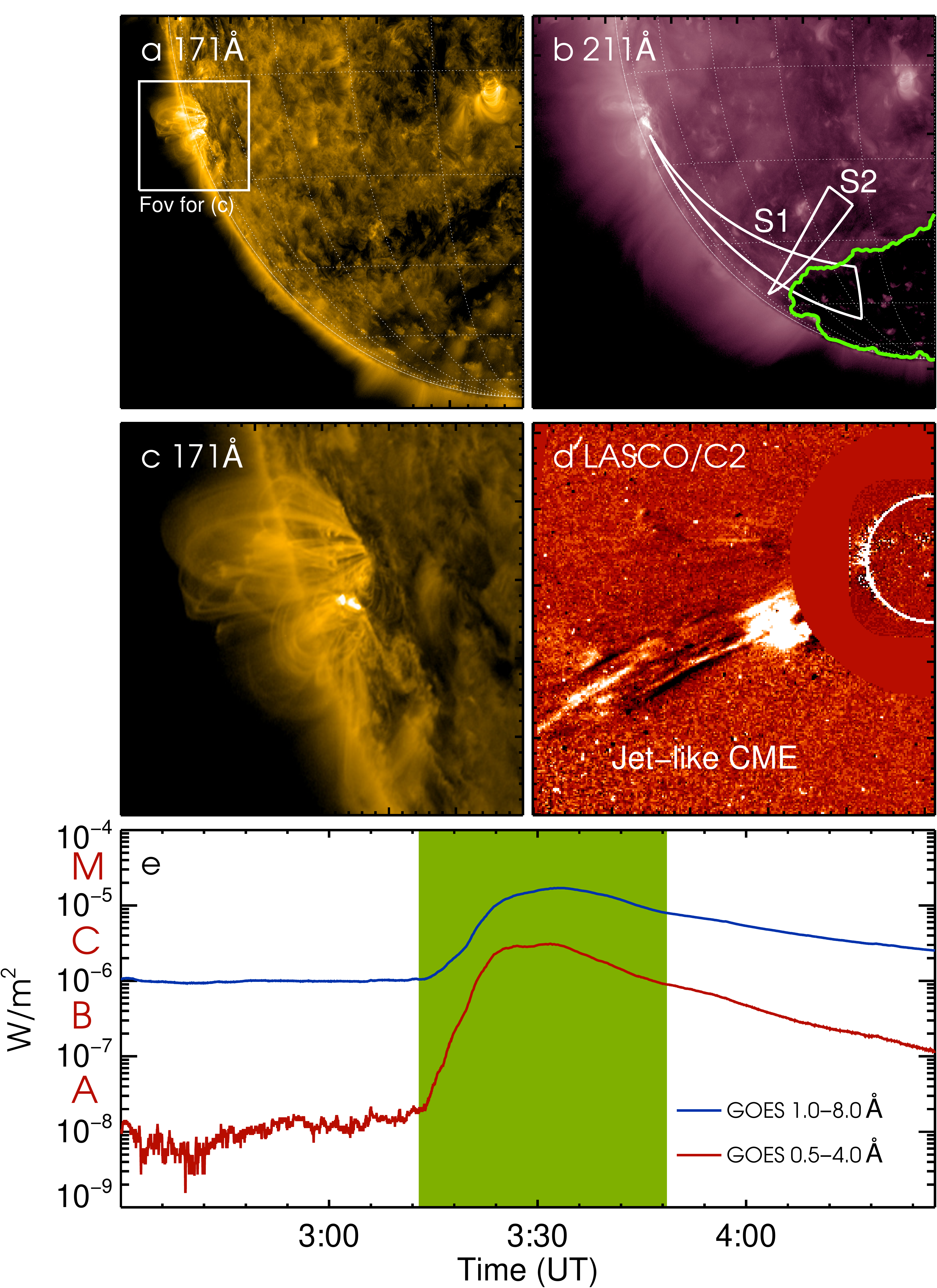}
    \caption{Overview of the event. Panels (a), (b) and (d) are AIA 171 \AA\, AIA 211 \AA\ and LASCO/C2 snapshots, respectively. The FOV of  the panel (c) is marked by the white box in panel (a). The incident and reflected wave trains are tracked with the two sectors indicated as S1 and S2 and the green line mark the location of the CH boundary in panel (b). Panel (e) shows the solar soft X-ray recorded by the GOES in 0.5--4.0 (red) and 1.0--8.0 (blue) \AA. The green region marks the start and end time of the flare in NOAA AR 12473.
}
    \label{overview}
\end{figure}

\begin{figure}
\centering
\includegraphics[width=8.5cm]{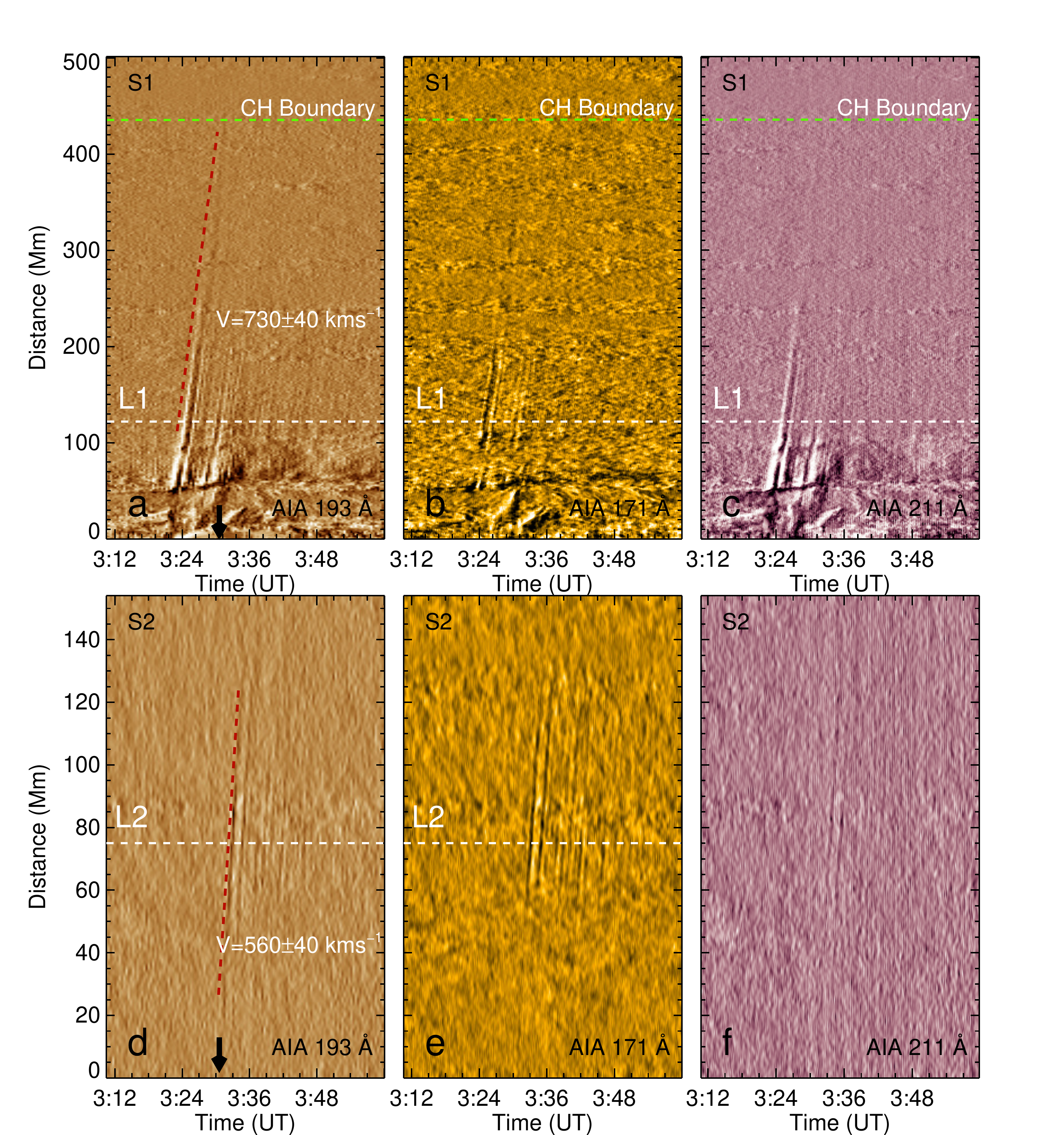}
\caption{
The top and bottom rows show the time-distance stack plots obtained from AIA 193 \AA, 171 \AA, and 211\AA\ running-difference images along the sectors S1 and S2 as shown in Fig.~\ref{overview}(b), respectively. The horizontal dashed lines indicate the positions for extracting signals for wavelet analysis, whereas the oblique red dotted lines depict the linear fitting results for estimating the speeds of wavefronts. The green dashed line in the top panels represents the boundary of the southern polar CH.}
\label{slice}
\end{figure}

To reveal the speeds of the wave train, we selected the sectors S1 and S2 as shown in Fig.~\ref{overview}(b) in the AIA 211 \AA\ channels. Sectors S1 (width 20$^\circ$) and S2 (width 8$^\circ$) were placed respectively over the incident and reflected wave trains' tracks, and they are selected in almost the exact directions of the corresponding wave trains. The time-distance stack plots are displayed in Fig.~\ref{slice}, in which one can identify that the incident and reflected wave train propagated at speeds of \speed{730} and \speed{560}, respectively. Obviously, the wave train’s speeds are close to that of classical EUV waves whereas significantly lower than that of the QFP wave train (around \speed{ 2000} reported by \cite{2011ApJ...736L..13L}). This may reflect the difference of the fast magnetosonic speed in the quiet-Sun (as shown in the present paper) and in funnel-like magnetic structures stemming from active regions (as reported by \cite{2011ApJ...736L..13L}). Therefore, the wave train under investigation in this paper should be classified as a broad wave train, according to the classification of \cite{shen2021sorev}. We estimated the error from making the fit ten times by placing two points along the front and deriving the average velocity \citep{2012ApJ...756..143O}. It is worth noting that only the first strongest wavefront can be tracked from the erupting origin to the CH boundary, and no significant wave signal that the wavefront intrudes into the CH as reported by \citep{2006ApJ...647.1466V}. The first wavefront encountered the CH boundary at about 3:32 UT (see the black arrow in Fig.~\ref{slice} (a)). Almost at the same time, the first reflected wavefront appeared (see the black arrow in Fig.~\ref{slice}(e)), followed by a series of wavefronts at the time-distance stack plot. The trailing wavefronts are so weak that it is not discernable from the time-distance stack plots in Fig.~\ref{slice} after propagation of about 200 Mm but can be discerned in the animation. From the time-distance stack plots, there is no significant signal of a reflected wave train in the AIA 211 \AA\ channel, indicating the energy dissipation of the wave train as its impingement and long-distance propagation.

\begin{figure*}
\centering
\includegraphics[width=15cm]{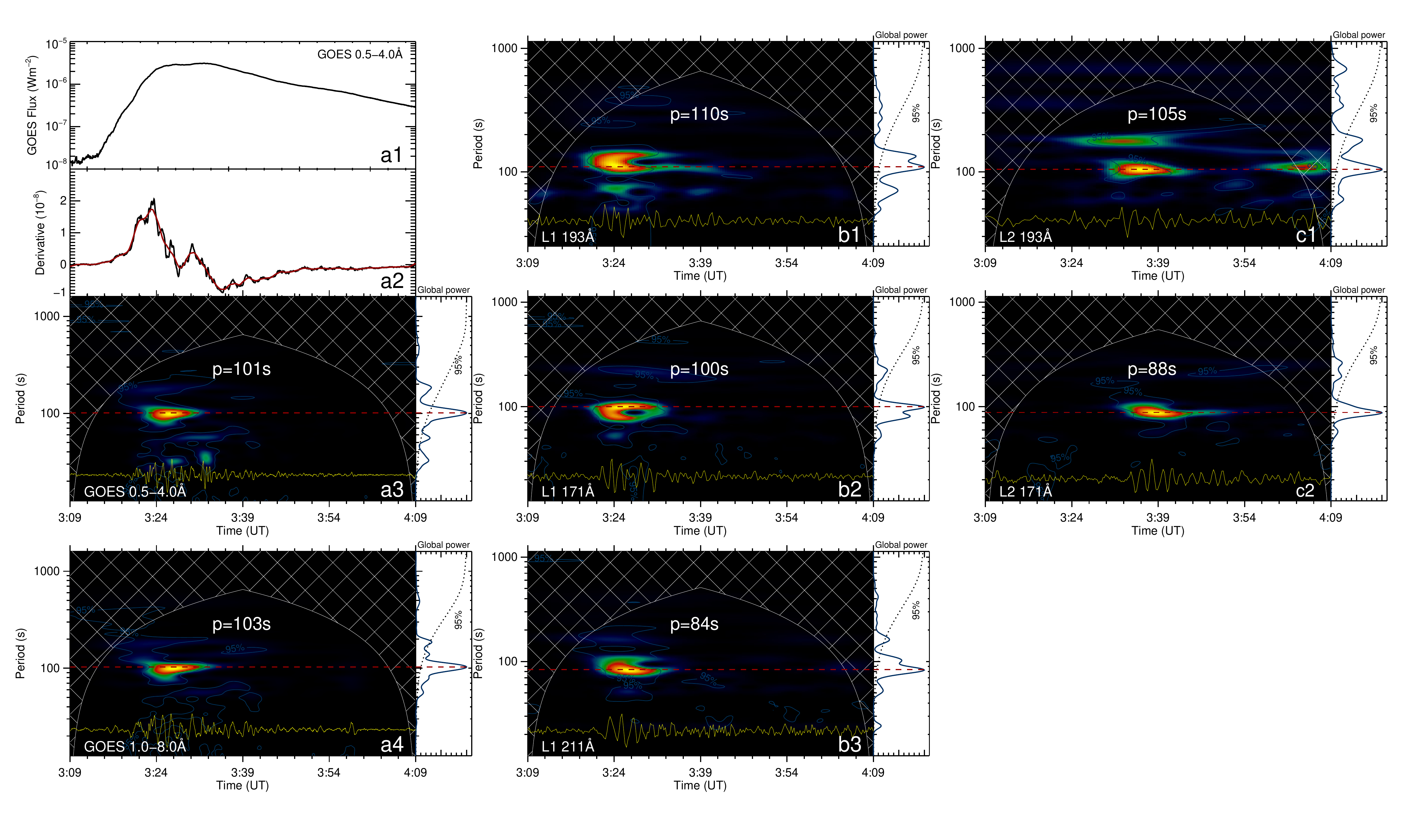}
\caption{
Periodicity analysis for the wave train and the associated flare pulsation. Panels (a1) and (a2) show the GOES 0.5--4.0 \AA\ soft X-ray flux and its temporal derivative (black curve in (a2)), respectively. The detrended intensity profile (yellow curve in panel (a3)) is obtained from the temporal derivative in (a2) by subtracting its smoothed intensity profile (red curve in (a2)). The smooth/detrended temporal derivative signal of the GOES 1.0--8.0 \AA\ soft X-ray flux is overlaid in panel (a4), while the yellow curves in panel (b1)--(c2) are smooth/detrended intensity profiles of the horizontal white dashed lines labeled ``L1'' and ``L2'' in Fig.~\ref{slice}, and their corresponding wave wavelet power spectrums are displayed in each panel. The cross-hatched regions in panels (a3)-(c2) indicate the cone of influence region due to the edge effect of the data, while the dotted and red horizontal dashed lines separately mark the 95\% confidence level and the period, respectively.
}
\label{wavelet}
\end{figure*}

To analyze the origin of the  EUV wave train, we applied the Morlet wavelet analysis \citep{1998BAMS...79...61T} to analyze the periodicities of the associated impulsive M 1.6 flare and the wave train. Since RHESSI had a data gap during the flare, thus temporal-derivative of GOES soft X-ray 0.5--4.0 \AA\ and 1.0--8.0 \AA\ fluxes as a proxy of the corresponding hard X-rays are used to analyze the periodicity of the flare according to the Neupert effect \citep{1968ApJ...153L..59N,2005ApJ...621..482V,2010ApJ...717.1232N}. Fig.~\ref{wavelet}(a1) displays GOES 0.5--4.0 \AA\ soft X-ray flux. Its temporal-derivative curve (black) is shown in Fig.~\ref{wavelet}(a2). After subtracting a 100 seconds smoothed curve (red) of the temporal-derivative curve, we get the detrended signal, and the result is overlaid in Fig.~\ref{wavelet}(a3) (see the yellow curve). Using the same method, we obtain the detrended, temporal-derivative curve of GOES 1.0--8.0 \AA\, and the result is overlaid in Fig.~\ref{wavelet}(a4). Using such a detrended temporal-derivative curve as input, we could see that the period of flare is about 100 seconds (see Fig.~\ref{wavelet}(a3)--(a4)).

|
For the periodicities of the wave train, we analyzed the intensity profile extracted along the white horizontal lines marked ``L1'' and ``L2'' as shown in Fig.~\ref{slice}. Using its detrended intensity profiles as input, we obtained that the average periods of the incident and reflected wave trains are 84--110 seconds and 88--105 seconds (average values are 98 seconds and 97 seconds), respectively (see Fig.~\ref{wavelet}(c1) and (c2)). This result satisfies the characteristics of a true wave, i.e., the frequency of wave trains does not change during propagating process or interaction with other structures.  From the wavelet analysis results, one can find that the period of the wave train is similar to that of the pulsations in the flare, suggesting that these wave train  were possibly initiated by the quasi-periodic pressure pulses launched by the intermittent energy-releasing process in the flare. Interestingly, a sub-period remains seen in Fig.~\ref{wavelet} b1-b3 and c1 for both incident  and reflected wave train. The sub-period for the incident and reflected wave trains from Fig~\ref{wavelet} b1-b3 and c1 are 132, 80, 103, and 180 seconds. In addition, a typical feature of the signature is a bifurcation and an inverse tadpole ( cf, \cite{1983Natur.305..688R,1984ApJ...279..857R,2004MNRAS.349..705N} for the tadpole in wavelet spectrum) in the wavelet spectrums, especially in Fig~\ref{wavelet} b1-b3, and c2. This special shape wavelet spectrum will be discussed in the next paper.

\subsection{Total Reflection of the Quasi-Periodic EUV Wave train}
Like any true wave, the EUV waves will exhibit various behaviors when encountering the region of suddenly reduced plasma density. In addition to the total reflection, all other true wave features, such as partial reflections, refraction, and transmissions, have been observed so far. This section will evidence that the wave train was totally reflected by the CH boundary.

We moved and projected S1 and S2 on the disk center, considering that the projection effect on this limb event, as shown in Fig.~\ref{dem}(a). Combining the incidence angle equals the reflected angle (i.e., $\theta_i=\theta_r$) and the angle between the incident and reflected wave trains (two red axes of the sectors S1 and S2) is about 115$^\circ$, we arrive at the incidence angle, $\theta_i $, is about 33$^\circ$, as shown in the inset in the left upper corner of Fig.~\ref{dem}(a). According to \cite{2021A&A...651A..67P}, the critical angle, $\theta_c$, depends only on the density contrast, i.e., $\theta_c=\cos^{-1}(\sqrt{\rho_c})$, in which the $\rho_c=\rho_i/\rho_o$ is the density contrast of inside and outside CH. To estimate the density contrast, we follow the method provided by \cite{2020SoPh..295....6S} to perform DEM analysis for the CH, and one of the examples is shown Fig.~\ref{dem}(b1). All of the AIA images used for DEM were binned by $8\times8$ pixels to enhance the signal-noise ratio and then further deconvolved with the instrument Point Spread Function (PSF) provided by the \textsf{aia\_calc\_psf.pro} routine to reduce the effect of instrument stray light. The plasma density can be derived from the formula $\bar{\rho}=\sqrt{\frac{\rm EM}{h}}$, where EM is the integration result of the DEM over the temperature range, and $h$ is the column height (taken here as 42 Mm and 90 Mm for CH and quiet-Sun region respectively, cf. \cite{2020SoPh..295....6S, 2019ApJ...882...90L}) of emitting plasma along the line of sight . We selected ten successive segments along the CH boundary near the waves-CH interaction location and estimated the densities in the patches outward and inward of the CH boundary of each individual segment. Fig.~\ref{dem} (b2) shows that the plasma density inside and outside of the CH boundary are in the range of $1.5-2.0\times10^8$cm$^-3$ and $2.5-3.0\times10^8$cm$^-3$, respectively, which are slightly lower than that reported by \cite{2020SoPh..295....6S} through studying the low-latitude CHs. This slight divergence should be the result of the projection effect. The density contrast for ten segments inside and outside CH are plotted in Fig.~\ref{dem}(b3). From the density contrast distribution, one can find that the average density contrast $\bar{\rho_c}$ is about 0.62. Based on the critical angle $\theta_c=\cos^{-1}(\sqrt{\rho_c})$ \citep{2021A&A...651A..67P}, we get the critical angle $\theta_c$ is about $38^\circ$, which is larger than the incidence angle $\theta_i$ ($33^\circ$), suggesting that the incident wave train was totally reflected at the CH boundary, resulting in the secondary wave train. According to the theoretical analysis by \cite{2021A&A...651A..67P}, the phase inversion angle is defined as $\theta_p=\rm {cos}^{-1}\lbrack\sqrt{\frac{\rho_c+\rho_c^2}{1+\rho_c^2}}\rbrack$. Here, we take the density contrast $\rho_c = 0.62$, we reach a phase inversion angle $\theta_p =31^\circ$, which is in the range of $0 <\theta_i<\theta_p$. This result further confirms that the incident wave train is total reflected at the CH boundary.

\section{Discussion and Conclusions}

Using high spatio-temporal resolution imaging observations from AIA/SDO, we have presented a rare observation of a series of wavefronts interaction with the southern polar CH that occurred on 2015 December 22 in NOAA AR 12473. This event is associated with an M1.6 flare and a jet-like CME. The wave train launched from the flare kernel, propagated along the solar surface, finally reflected at the CH boundary. The speeds of the incident and reflected wave trains are \speed{730} and \speed{560}, respectively. Their period is about 100 seconds, which is similar to that of the pulsation flare. Thus we propose that the wave train was initiated by the intermittent energy-release process in the flare, at least in this event. The observed evolutionary process together with the fact that the incidence angle is less than the critical angle (i.e., $\theta_i < \theta_c$), suggests that the flare-initiated, quasi-periodic wave train was totally reflected at the CH boundary.

\begin{figure*}
\centering
\includegraphics[width=15cm]{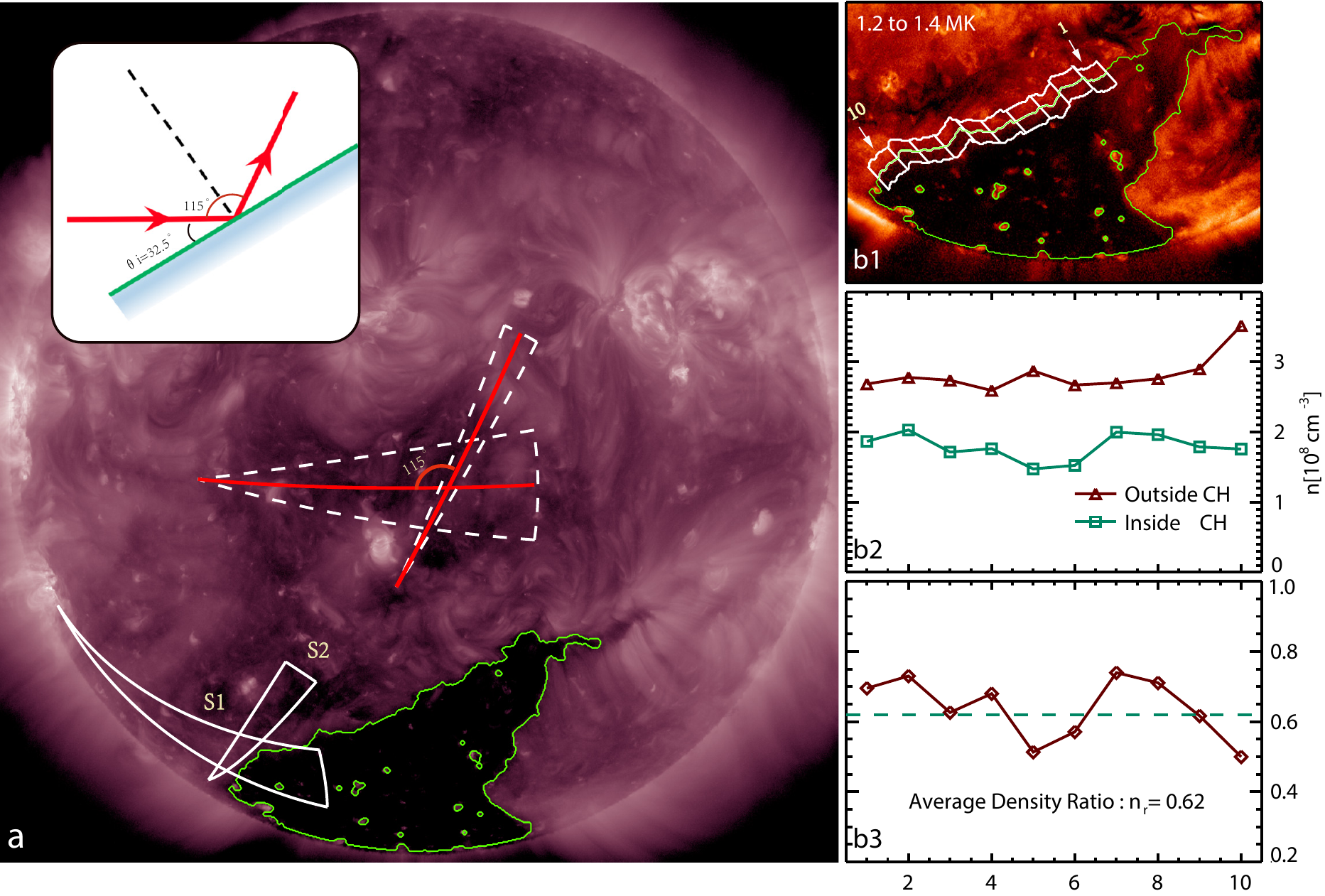}
\caption{
The geometrical properties of the wave trains and the density contrast surrounding the CH boundary. (a) AIA 211\AA\ full-disk images of green contour represent the CH boundary extracted with the CATCH, and the sectors located on the central disk show the S1 and S2 translation and projection results. (b1) shows the plasma total emission measure (EM) at the 1.2-1.4 MK temperature range. Twenty patches were selected to estimate the density contrast surrounding the CH boundary, of which ten are inside the CH, and ten are outside the CH. The green and red lines in panel (b2) show the density inside and outside of the CH, respectively. (b3) shows the plasma density contrast over the inside and outside CH regions.}
\label{dem}
\end{figure*}

It is worth noting that most of the studies revealed that the EUV waves are tightly associated with CMEs,  whereas the relationship with flares is very weak \citep[e.g.,][]{2008ApJ...681L.113V,2010ApJ...716L..57V,2018ApJ...864L..24L,2021ApJ...911..118D}. At the same time, a considerable number of observational evidence show that these waves always exhibit a single diffuse wavefront. Another feature is the appearance of multiple wavefronts in a single eruption, which, however, had rarely been observed. According to the statistics by \cite{shen2021sorev}, there are only three EUV wave trains have been published. These wave trains with multiple wavefronts, such as one propagating along the solar surface in the current event, are significantly different from the narrow QFP magnetosonic wave trains \citep{2011ApJ...736L..13L}. The main dominant difference between the broad and narrow trains is their propagating medium: the former runs along the solar surface, whereas the latter is constrained in the closed- or open-loops.
The narrow QFP wave trains are mainly detected in the AIA 171 \AA\ and sometimes 193 \AA\ wavelengths. In contrast, the quasi-periodic EUV wave trains like the classical EUV waves are visible in AIA 171 \AA\, 193 \AA\, and 211 \AA\ wavelengths, suggesting different temperature ranges. More detailed information about these two types of wave trains we can find in a recent review paper \citep[][and references therein]{shen2021sorev}. These difference indicate that the observed wave train in the current event is a rare quasi-periodic wave train. In the scenario of a piston-driven shock, the wavefront should decouple from the driver, resulting in a single wavefront propagating freely on the solar surface. The single erupting jet-like CME as the driver can not excite multiple successive wavefronts, indicating that the observed EUV wave train was unlikely launched by the jet-like CME.  Since the similar period of the flare pulsations and the wave train, we incline to the notion that it was initiated by the quasi-periodic pressure pulses launched by the intermittent energy-releasing process in the flare. As we have already mentioned, EUV waves, characterized by a single wavefront, are usually initiated by the CME and generally have a weak connection with flare. The present study showed strong evidence that EUV waves can also be excited impulsively by the quasi-periodicity of flare and appear multiple wavefronts. Recently, \cite{2021SoPh..296..169Z} reported a similar example, i.e., multiple consecutive wavefronts were propagating continuously inside a CME bubble that had already completed developmental completion. This event is also hard to explain using the CME expansion.  The appearance of the first wavefront in present event has a delay of several minutes compared with the onset time of the flare. A possible explanation is that the wave needs time to build up a large amplitude or shock to be observed. 

Observational studies of the interaction of the EUV waves with the CHs are scarce, but the appearance of the secondary waves is found to be a pretty frequent phenomenon when the waves interact with CHs or ARs \citep{ 2000ApJ...543L..89W,2009ApJ...691L.123G,2019ApJ...870...15L}. These observational findings were verified by numerical simulation \citep{ 2017ApJ...850...88P,2018ApJ...857..130P, 2018ApJ...860...24P,2018A&A...614A.139A}. In accordance with these observational and simulation wave-like features confidently confirm the interpretation that it is a fast-mode magnetohydrodynamic (MHD) wave. However, as we have already mentioned, the total reflected wave train is a basic behavior of waves, has hardly ever been observed in the solar corona. Studying the wave-CH interaction and its secondary waves can provide much information about the CHs themselves, particularly their boundaries. Because of that, (i) the observed CH areas are well correlated with the solar wind speed measured in situ at 1 AU, which is of crucial importance parameters for Space Weather forecasting models. (ii), accurate extraction of the boundaries of the CHs is the premise of investigating the photospheric magnetic field encompassed by the CH region, which strongly correlated with the CH area \citep{2018ApJ...861..151H,2018ApJ...863...29H}. As depicted in the Fig.~\ref{dem} inset, the incidence angle is about $33^{\circ}$. Using the method provided by \cite{2021A&A...651A..67P}, we get the critical angle of the wave train was about $38^\circ$, which is significantly large than the incidence angle, providing strong evidence that the incident wave train was indeed totally reflected by the southern polar CH. Recently, based on a few statistical studies, it has been shown that the density contrast to the CH boundary is in the range of 0.1-0.6 \citep{2020SoPh..295....6S,2019SoPh..294..144H} and the corresponding critical angle $\theta_c$ is in the range of $39^\circ - 72^\circ$ \citep{2021A&A...651A..67P}. This more representative statistical result further proves that the wave train is totally reflected at the CH boundary since it is larger than the incidence angle $\theta_i$ in the current event. 

The wave-CH interactions allow us to study the influence of the CH on wave parameters, such as the velocity and the amplitude \citep{2010ApJ...708.1639M,2012ApJ...756..143O,2018ApJ...864L..24L,2019ApJ...878..106H}. The observed secondary wave signals are usually weak, resulting in the parameters of the measured secondary waves being relatively rough and of low quality. Therefore, studying more clearly captured examples of wave-CH interactions can help us deepen our understanding of the influence of coronal holes on coronal waves and get more information about the CHs.

%================================

\begin{acknowledgements} 
We want to thank the anonymous referee for his/her many valuable suggestions and comments for improving the quality of this article. We also thank Prof. Dr. Astrid M. Veronig and Dr. Stephan G. Heinemann for data processing and useful suggestions. Moreover, the authors want to acknowledge SDO/AIA, GOES, and SOHO/LASCO science teams for providing the data.  This work is supported by the Natural Science Foundation of China (12173083,11922307,11773068,11633008), the Yunnan Science Foundation for Distinguished Young Scholars (202101AV070004), the National Key R\&D Program of China (2019YFA0405000),  and the West Light Foundation of Chinese Academy of Sciences.
\end{acknowledgements}

\end{document}